\documentclass[aps,prd,twocolumn,groupedaddress,preprintnumbers,showpacs]{revtex4}

\usepackage{amsmath,amsfonts,amssymb,mathrsfs,graphicx}

\newcommand{\ie}{{\it i.e.}}

\newcommand{\eg}{{\it e.g.}}

\newcommand{\cf}{{\it cf.}}

\newcommand{\eq}{Eq.}
\newcommand{\eqs}{Eqs.}
\newcommand{\fig}{Fig.}

\newcommand{\Ref}{Ref.}
\newcommand{\Refs}{Refs.}

\begin{document}

\preprint{hep-ph/0112226}
\preprint{TUM-HEP-450/01}

\title{Tests of CPT Invariance at Neutrino Factories}

\author{Samoil M. Bilenky}
\email[E-mail address: ]{sbilenky@ph.tum.de}
\altaffiliation{On leave from Joint Institute for Nuclear Research,
Dubna, Russia.}
\affiliation{Institut f{\"u}r Theoretische Physik, Physik-Department,
Technische Universit{\"a}t M{\"u}nchen, James-Franck-Stra\ss{}e,
85748 Garching bei M{\"u}nchen, Germany}

\author{Martin Freund}
\email[E-mail address: ]{mfreund@ph.tum.de}
\affiliation{Institut f{\"u}r Theoretische Physik, Physik-Department,
Technische Universit{\"a}t M{\"u}nchen, James-Franck-Stra\ss{}e,
85748 Garching bei M{\"u}nchen, Germany}

\author{Manfred Lindner}
\email[E-mail address: ]{lindner@ph.tum.de}
\affiliation{Institut f{\"u}r Theoretische Physik, Physik-Department,
Technische Universit{\"a}t M{\"u}nchen, James-Franck-Stra\ss{}e,
85748 Garching bei M{\"u}nchen, Germany}

\author{Tommy Ohlsson}
\email[E-mail address: ]{tohlsson@ph.tum.de}
\affiliation{Institut f{\"u}r Theoretische Physik, Physik-Department,
Technische Universit{\"a}t M{\"u}nchen, James-Franck-Stra\ss{}e,
85748 Garching bei M{\"u}nchen, Germany}

\author{Walter Winter}
\email[E-mail address: ]{wwinter@ph.tum.de}
\affiliation{Institut f{\"u}r Theoretische Physik, Physik-Department,
Technische Universit{\"a}t M{\"u}nchen, James-Franck-Stra\ss{}e,
85748 Garching bei M{\"u}nchen, Germany}

\date{\today}

\begin{abstract}
We investigate possible tests of CPT invariance on the level of event rates at
neutrino factories. We do not assume any specific model but phenomenological
differences in the neutrino-antineutrino masses and mixing angles in a Lorentz
invariance preserving context, such as it could be induced by physics beyond
the Standard Model. We especially focus on the muon neutrino and
antineutrino disappearance channels in order to obtain constraints on
the neutrino-antineutrino mass and mixing angle differences; we found,
for example, that the sensitivity $|m_3 - \overline{m}_3| \lesssim 1.9
\cdot 10^{-4} \, \mathrm{eV}$ could be achieved.
\end{abstract}

\maketitle

\section{Introduction}
\label{sec:intro}

The CPT theorem \cite{CPT} is one of the milestones of local quantum
field theory. It is based on such general principles as Lorentz
invariance, the connection of spin and statistics, and the locality and
hermiticity of the Lagrangian. The ${\rm SU(3)} \times
{\rm SU(2)} \times {\rm U(1)}$ Standard Model of Elementary
Particle Physics (SM), for which the CPT theorem is valid, is in very
good agreement with all existing experimental data.
Beyond the SM, like in string theory models or in models involving extra
dimensions, CPT invariance could be violated \cite{Kost1,Kost2}. Thus, the
search for possible effects of CPT violation is connected to the
search for physics beyond the SM. Many different tests of CPT
invariance have been carried out. So far, no CPT violation has been
found and rather strong bounds on the corresponding parameters have
been obtained \cite{Groom:2000in}.

One of the basic consequences of the CPT
theorem is the equality between the masses of particles and their corresponding
antiparticles. A strong bound on a
possible violation of CPT invariance has been obtained from the
$K^0$-${\bar K}^0$ system. This violation is characterized by the parameter
\begin{equation} 
\Delta \equiv \frac{{\cal H}_{{\bar K^0};{\bar K^0}} -
{\cal H}_{K^0;K^0}}{2(\lambda_{L}-\lambda_{S})},
\label{eq:delta}
\end{equation}
which can be related to measurable quantities \cite{Angelopoulos:1998dw}.
In \eq~(\ref{eq:delta}), $\lambda_{L,S} \equiv m_{L,S} - \frac{\rm{i}}{2}
\Gamma_{L,S}$, $m_{L,S}$ and $\Gamma_{L,S}$ are the masses and the
total decay widths of the $K^{0}_{L}$ and $K^{0}_{S}$ mesons, respectively, and
${\cal H}$ is the effective non-Hermitian Hamiltonian of the $K^0$-${\bar
K}^0$ system in the representation $| K^0 \rangle$ and $| {\bar K}^0
\rangle$, which are eigenstates of the Hamiltonian of strong and
electromagnetic interactions. For the complex diagonal matrix
elements, we have ${\cal H}_{{\bar K}^0;{\bar K}^0} = m_{{\bar K^0}} -
\frac{\rm i}{2} \Gamma_{{\bar K}^0}$ and ${\cal H}_{K^0;K^0} = m_{K^0}
- \frac{\rm i}{2} \Gamma_{K^0}$, where $m_{K^0,{\bar K}^0}$ and
$\Gamma_{K^0,{\bar K}^0}$ are the bare masses and the total decay
widths of the $K^0$ and ${\bar K}^0$ mesons, respectively, with
corrections due to weak interactions. The CPLEAR experiment obtained 
\cite{Zavrtanik:2001zq} $$ |m_{K^0} - m_{{\bar K}^0}| = (-1.5 \pm 2) \cdot
10^{-18} \, \rm{GeV}. $$
Using all relevant data on the $K^0$-$\bar K^0$ system, it follows that
\cite{Groom:2000in}
$$
\frac{|m_{K^0} - m_{{\bar K}^0}|}{m_{\rm average}} \lesssim 10^{-18},
$$
where $m_{\rm average} \equiv (m_{K^0} + m_{\bar K^0})/2$.
Recently, also an upper bound on the mass difference between the
$B^0_d$ and ${\bar B}^0_d$ mesons has been obtained \cite{Leonidopoulos:2001ci}
$$
\frac{|m_{B^0_d} - m_{{\bar B}^0_d}|}{m_{B^0_d}} \lesssim 1.6 \cdot 10^{-14}.
$$

Here, we will consider possible CPT invariance tests that can be
performed in future high-precision experiments with neutrinos
from neutrino factories, which are now under active investigation
\cite{Geer:1998iz,Blondel:2000gj,Barger:1999jj,Freund:1999,Freund:2001ui,Cervera:2000kp,Albright:2000xi,Adams:2001tv}.
They have mainly been proposed to study neutrino
oscillations in detail. In addition, in the framework of Lorentz non-invariant
models, possible CPT invariance tests with neutrino experiments
have been discussed (see, \eg, \Refs~\cite{Coleman:1998ti,Barger:2000iv}). 

Compelling evidence for neutrino oscillations
has been found by atmospheric \cite{atmosS-K} and solar
\cite{solar,solarS-K,Ahmad:2001an} neutrino experiments.
The following best-fit value for the atmospheric mass squared
difference $\Delta m^2_{\rm atm}$ has been obtained \cite{atmosS-K}:
$$
\Delta m^2_{\rm atm} \simeq 2.5 \cdot 10^{-3} \, {\rm eV}^{2}.
$$
{}From the global analysis of all solar neutrino data, several allowed
regions in the neutrino oscillation parameter space
have been found. For the preferred so-called large mixing angle
(LMA) solution in \Ref~\cite{Bahcall:2001zu}, the solar mass squared
difference has been determined to be
$$
\Delta m^2_\odot \simeq 4.5 \cdot 10^{-5} \, {\rm eV}^{2}.
$$
Furthermore, there are at present indications for neutrino
oscillations with an even larger mass squared difference, which were
found by the LSND experiment \cite{LSND}. {}From the analysis of the
data of the LSND experiment, the best-fit value of the neutrino mass
squared difference \cite{stancu}
$$
\Delta m^{2}_{\rm LSND} \simeq 0.24 \, \rm{eV}^{2}
$$
was found.

The strongest kinematical bound on the absolute neutrino mass scale $m_1$ is
obtained from the endpoint of the $\beta$-spectrum of $^3{\rm H}$. The latest
measurements yielded $m_{1} \lesssim 2.2 \, {\rm eV}$ \cite{MAINZ,TROITSK}.
{}From neutrinoless double $\beta$-decay there exists also a strong
bound $| \langle m \rangle | \equiv \left| \sum_i U_{ei}^2 m_i \right|
\leq (0.2 - 0.6) \, \mathrm{eV}$ for Majorana masses (for an overview see, \eg,
\Ref~\cite{Law:2001sy}). Here $U_{ei}$ are matrix elements of the
neutrino mixing matrix $U$ and $m_i$ are the masses of the neutrino
mass eigenstates. Furthermore, somewhat weaker but similar bounds emerge from
astrophysics and cosmology. It nevertheless follows from the existing neutrino
data that neutrino masses are not equal to zero and that they are much smaller
than the masses of all other fundamental fermions (leptons and quarks). From
empirical lepton and quark mass patterns a hierarchical (or inverse
hierarchical) mass pattern seems to be rather plausible \cite{Lindner:2001kd}.

It is a general belief that the smallness of the
neutrino masses requires some new mechanism beyond the SM.
The classical mechanism of neutrino mass generation is the see-saw
mechanism \cite{seesaw}, which connects the smallness of the neutrino
masses with the violation of lepton numbers at an energy scale much higher
than the electroweak scale. In
this case, massive neutrinos have to be Majorana particles and the neutrino
masses have to satisfy a hierarchy relation. The see-saw mechanism is
based on local quantum field theory, and therefore, violation of CPT
invariance cannot be expected.

Furthermore, it has recently been suggested \cite{Arkani-Hamed:1998vp} that the
smallness of the neutrino 
masses could have a natural explanation in models with large extra spatial 
dimensions. In such models, the smallness of the Dirac neutrino masses
follows from the suppression of Yukawa interactions of the 
left-handed neutrino fields, localized on a three-dimensional brane,
and the singlet right-handed neutrino fields propagating together
with the gravitational field in a bulk. In models with $n$ extra
dimensions, the neutrino masses are proportional to 
$$
\sqrt{\frac{1}{M^n V_n}} = \frac{M}{M_G} \simeq 10^{-16}
\frac{M}{\rm{TeV}},
$$
where $V_n$ is the volume of the extra space, $M_G \simeq 1.2 \cdot
10^{19} \, \rm{GeV}$ is the Planck mass, and $M \simeq 1 \, \rm{TeV}$
is the Planck mass in the $4+n$ dimensional space. Moreover, there are
other approaches to the generation of small Dirac or Majorana
neutrino masses in models with extra dimensions (see,
\eg, \Refs~\cite{Grossman:1999ra,Huber:2001ug}). 
Since the symmetries of the SM are violated in the bulk, neutrino mass
generation in extra dimension models is a plausible candidate for the
violation of CPT invariance \cite{Barenboim:2001ac}.

In order to accommodate all existing neutrino oscillation data,
including the data of the LSND experiment, it is necessary to have
three independent mass squared differences. Thus, we need to assume that
there exist (at least) four massive mixed neutrinos, \ie, in addition to
the three active flavors $\nu_e$, $\nu_\mu$, and $\nu_\tau$ at
least one sterile neutrino has to exist \cite{Bilenkii:1998dt}.

In \Refs~\cite{Murayama:2000hm,Barenboim:2001ac}, it was assumed that
CPT violation in the neutrino sector can be so strong that the mass spectra
of neutrinos $\nu_i$ and antineutrinos $\overline{\nu_i}$ are
completely different. In this case, it is possible to describe atmospheric,
solar, and LSND neutrino data with a framework of three massive
neutrinos and three massive antineutrinos (assuming that $\Delta
m_{\rm LSND}^2$ belongs to the antineutrino spectrum). Such an extreme
picture can, in principle, be tested by the future MiniBooNE
\cite{Bazarko:2000id}, KamLAND \cite{Piepke:2001tg}, and other similar
neutrino experiments \cite{Barenboim:2001ac}.

In \Ref~\cite{Barger:2000iv}, the effect of a term in the neutrino
Hamiltonian violating CPT and Lorentz invariance has been
considered and the $\nu_\mu \to \nu_\mu$ and $\overline{\nu_\mu} \to
\overline{\nu_\mu}$ transition probabilities with the $\nu_\mu$ and
$\overline{\nu_\mu}$ coming from neutrino factories have been
calculated. It was demonstrated that in such a model the effects of CPT
violation could be rather large in a wide range of the
corresponding parameter values.

\section{Basic formalism}
\label{sec:bf}

In this paper, we will assume Lorentz invariance and consider
possible violation of CPT invariance by the mechanism of neutrino mass
generation. In the case of the usual neutrino
mixing, we have \begin{equation}
\nu_{\alpha L} = \sum_i U_{\alpha i} \, \nu_{iL},
\label{eq:nufield}
\end{equation}
where $U$ is a unitary mixing matrix and $\nu_{i}$ are the neutrino fields
(Dirac or Majorana) with masses $m_{i}$. The neutrino flavor state
$|\nu_\alpha\rangle$ is given by
\begin{equation}
|\nu_\alpha\rangle = \sum_i U_{{\alpha}i}^* \; |\nu_i; m_i,L\rangle,
\label{eq:nustate}
\end{equation}
where $|\nu_i;m_i,L\rangle$ are the neutrino states with masses $m_i$,
negative helicity $L$, 3-momentum ${\bf p}$, and energy
\begin{equation}
E_i = \sqrt{m_i^2 + {\bf p}^2} \simeq p + \frac{m_i^2}{2 p}
\end{equation}
in the ultra-relativistic limit.~\footnote{In \eq~(\ref{eq:nufield}),
the mixing matrix elements $U_{\alpha i}$ are defined for neutrino
fields $\nu_{\alpha L}$ in coordinate space. Neutrino flavor states $|
\nu_\alpha \rangle$ (in momentum space) are, however, created from the
vacuum by creation operators of definite momentum. The transition from
coordinate space to momentum space explains why the $U_{\alpha
i}^\ast$'s show up in \eq~(\ref{eq:nustate}).}
For the antineutrino flavor state $|\overline{\nu_{\alpha}}\rangle$ we have
\begin{equation}
|\overline{\nu_\alpha}\rangle = \sum_i U_{{\alpha}i} \;
|\overline{\nu_{i}}; m_i,R\rangle
\label{eq:antiDirac}
\end{equation}
in the case of Dirac neutrinos and
\begin{equation}
|\overline{\nu_\alpha}\rangle = \sum_i U_{{\alpha}i} \; |\nu_i;m_i,R\rangle,
\label{eq:antiMajorana}
\end{equation}
in the case of Majorana neutrinos. In these relations,
$|\overline{\nu_{i}};m_i,R\rangle$ and $|\nu_i;m_i,R\rangle$
are the right-handed antineutrino and Majorana neutrino states,
respectively, which also have the 3-momentum ${\bf p}$ and the energy
$E_{i}$.

Assuming the usual Lorentz invariant propagation of neutrino states for the
neutrino and antineutrino transition probabilities in vacuum, we find the
expressions
\begin{equation}
{\rm P}(\nu_\alpha\to\nu_{\alpha'}) = \left|\sum_i U_{{\alpha'} i} \; e^{-i
\Delta m^2_{i1} \frac{L}{2E}} \; U_{{\alpha}i}^* \right|^2
\label{eq:Paa'}
\end{equation}
and 
\begin{equation}
{\rm P}(\overline{\nu_\alpha}\to\overline{\nu_{\alpha'}}) =
\left| \sum_i U_{{\alpha'} i}^* \; e^{-i \Delta m^2_{i1} \frac
{L}{2E}} \; U_{{\alpha}i} \right|^2,
\label{eq:aPaa'}
\end{equation}
which automatically satisfy the relation
\begin{equation}
{\rm P}(\nu_\alpha\to\nu_{\alpha'})=
{\rm P}(\overline{\nu_{\alpha'}}\to\overline{\nu_{\alpha}}).
\label{eq:P=P}
\end{equation}
In \eqs~(\ref{eq:Paa'}) and (\ref{eq:aPaa'}), $\Delta m^2_{ij} \equiv
m^{2}_{i} - m^{2}_{j}$ is the mass squared difference, $L \simeq t$ is
the distance between the source and detector, and $E$ is the neutrino energy.
Note that \eq~(\ref{eq:P=P}) is a consequence of CPT invariance inherent
to standard neutrino mixing and oscillations.

If the generation mechanism of neutrino masses and mixings violates 
CPT invariance, then the relations for antineutrino
flavor states will differ from \eqs~(\ref{eq:antiDirac}) and
(\ref{eq:antiMajorana}).
In the case of massive Dirac neutrinos, the antineutrino
masses $\overline{m}_i$ will be different from the neutrino masses $m_i$,
and the mixing matrices will, in general, not be connected by complex
conjugation. Thus, for the antineutrino flavor states we have
\begin{equation}
|\overline{\nu_\alpha}\rangle = \sum_i \overline{U}_{{\alpha}i} \;
|\overline{\nu_i};\overline{m}_i,R\rangle.
\end{equation}
In the case of massive Majorana neutrinos, neutrinos and antineutrinos
are identical. For the right-handed antineutrino flavor states, we
therefore have
\begin{equation}
|\overline{\nu_\alpha}\rangle = \sum_i \overline U_{{\alpha}i} \;
|\nu_i; m_i,R\rangle.
\end{equation}
Further on, we will assume that there is no violation of Lorentz invariance 
in the propagation of massive neutrinos and antineutrinos.

\section{CPT Tests at Neutrino Factories}

In this section, we will investigate the sensitivity of future
high-precision neutrino oscillation experiments at neutrino
factories to neutrino-antineutrino mass and
mixing angle differences. Neutrino factories
\cite{Geer:1998iz,Blondel:2000gj} will allow to investigate
the phenomenon of neutrino oscillations, which has been observed by the
atmospheric and solar
neutrino experiments, with unprecedented accuracy. It will be
possible to determine the leading neutrino oscillation parameters
$\Delta m_{32}^2$ and $\sin^2 2 \theta_{23}$ governing the $\nu_\mu
\to \nu_\tau$ oscillations in the atmospheric region very
well. Depending on their values, it will also be possible to limit or
to measure the mixing angle $\theta_{13}$ to search for the connected
matter effects and to discriminate between a hierarchical neutrino
mass spectrum and a mass spectrum with reversed hierarchy. In the most
likely LMA case, the effects of CP violation in the lepton sector can
be studied. Details of neutrino factory phenomenology can be found in
\Refs~\cite{Barger:1999jj,Freund:1999,Freund:2001ui,Cervera:2000kp,Albright:2000xi,Adams:2001tv}.
As we will show below, because of
the high precision of neutrino factories, we can estimate the
sensitivity of experiments to the presumably small violations of CPT
invariance in the neutrino sector, being an unambiguous sign of new physics.

At neutrino factories neutrinos will be produced in muon decays
$\mu^+ \to e^+ \nu_e \overline{\nu_\mu}$ (or $\mu^- \to e^- \overline{\nu_e}
\nu_\mu$). The straightforward way to test CPT invariance at neutrino
factories would be to check the appearance relation 
${\rm P}(\nu_e \to \nu_\mu) = {\rm P}(\overline{\nu_\mu} \to
\overline{\nu_e})$ (or ${\rm P}(\overline{\nu_e} \to \overline{\nu_\mu}) =
{\rm P}(\nu_\mu \to \nu_e)$) with neutrinos from $\mu^+$ ($\mu^-$) decays.
However, such tests would require to measure the sign of the charge of the
produced lepton. The sign of a muon charge can be determined very reliably,
but measuring the sign of an electron (or positron) charge is a rather
challenging problem. The possibility to measure the electron (or positron)
charge with moderate efficiency with liquid argon detectors would not be
precise enough. 
Therefore, we consider a CPT invariance test in the $\nu_\mu$ and
$\overline{\nu_\mu}$ disappearance channels by checking the equality
$$ {\rm P}(\nu_\mu \to \nu_\mu) = {\rm P}(\overline{\nu_\mu} \to
\overline{\nu_\mu}).
$$
The $\nu_\mu$ and $\overline{\nu_\mu}$ disappearance channels have
several advantages:
\begin{enumerate}
\item The effect of neutrino oscillations in the atmospheric mass
squared difference region is large.
\item The matter effects are small.
\item There is no relevant background from the $\overline{\nu_e}$'s
($\nu_e$'s), which are accompanying the $\nu_\mu$'s ($\overline{\nu_\mu}$'s)
in the decays of the $\mu^-$'s ($\mu^+$'s).
\item
The event rates are high for obtaining good statistical information.
\end{enumerate}
We will only consider the possible violation of CPT invariance in
the $\nu_\mu \to \nu_\mu$ and $\overline{\nu_\mu} \to
\overline{\nu_\mu}$ oscillations. If CPT invariance is violated,
then these oscillations will be characterized by the leading parameters
$\Delta m^{2}_{32}$, $\sin^2 2 \theta_{23}$ and $\Delta \overline{m}^{2}_{32}$,
$\sin^2 2 \bar\theta_{23}$, respectively.

In \Ref~\cite{Freund:2001ui}, a comprehensive study of the accuracy of the
measurement of neutrino oscillation parameters in neutrino factory experiments
was performed. Our calculations will be based on this study. Since matter
effects give only small contributions to the $\nu_\mu$ and
$\overline{\nu_\mu}$ survival probabilities, uncertainties in the Earth matter
density profile are of little importance for the parameter measurements.
In \Ref~\cite{Freund:2001ui}, \fig~3, the relative statistical errors of the 
parameters $\delta \Delta m^{2}_{32}$ and $\delta \theta_{23}$,
determined by a general analysis including correlations, are plotted as
functions of the luminosity 
$$ {\cal L} = 2 \, N_\mu \, m_{\rm
kt}, 
$$ where $N_\mu$ is the number of stored muons per year and $m_{\rm kt}$
is the mass of the detector in kilotons.

\begin{figure}[ht!]
\caption{\label{fig:asymmetry} 
The sensitivitities $\delta a_{\rm CPT}$ and $\delta b_{\rm CPT}$ of
an estimate of the asymmetries $a_{\rm CPT}$ and $b_{\rm CPT}$ at a
neutrino factory as functions of the 
luminosity ${\cal L}$. The solid curve refers to the mass asymmetry 
$a_{\mathrm{CPT}}$ (hierarchical or inverse hierarchical
mass spectrum only) and the dashed curve to the mixing angle asymmetry
$b_{\mathrm{CPT}}$. The underlying calculations in \Ref~\cite{Freund:2001ui},
\fig~3, were performed with $50 \, \mathrm{GeV}$ muon energy and baselines of
$7000 \, \mathrm{km}$ ($\theta_{23}$) and $3000 \, \mathrm{km}$ ($\Delta
m_{32}^2$).}
\vspace{1cm}
\begin{center}
\includegraphics[angle=-90,width=8cm]{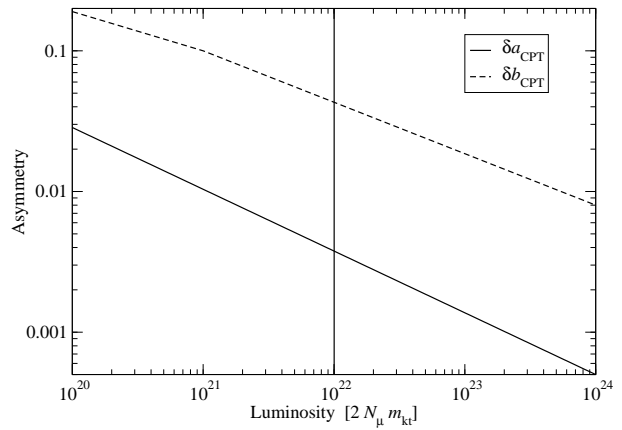}
\end{center} 
\end{figure}

Violation of CPT invariance in neutrino oscillations can be characterized
by the following parameters:
\begin{eqnarray}
\delta &\equiv& |\Delta m^2_{32} - \Delta \overline{m}^2_{32}|, \\
\epsilon &\equiv& | \sin^2 2 \theta_{23} - \sin^2 2 \bar\theta_{23}|.
\end{eqnarray}
If the minimal neutrino mass $m_1$ and the CPT violating effects
are small ($m_1 \ll \sqrt{\Delta m^{2}_\odot}$,
$|m_3 - \overline{m}_3|\ll {(m_{3})_{\rm average}}$), 
then we find for the hierarchical neutrino mass
spectrum or the spectrum with reversed hierarchy that
\begin{equation}
\delta \simeq 2 \, a_{\rm{CPT}} \, \Delta m^{2}_{32},
\label{eq:acptrel}
\end{equation}
where
\begin{equation}
a_{\rm{CPT}} \equiv \frac{|m_3 - \overline{m}_{3}|}{(m_{3})_{\rm average}}
\label{eq:acpt}
\end{equation}
is a dimensionless parameter which characterizes the violation of CPT
invariance. We can also write $\epsilon$ as
\begin{eqnarray}
\epsilon &\simeq& 2 \, b_{\rm{CPT}} \, \sqrt{\sin^2 2\theta_{23}} \,
\sqrt{1 - \sin^2 2\theta_{23}} \, \arcsin \sqrt{\sin^2 2\theta_{23}}
\nonumber\\
&=& 2 \, b_{\rm{CPT}} \, \theta_{23} \, \sin 4 \theta_{23}, \nonumber\\
\end{eqnarray}
where
\begin{equation}
b_{\rm{CPT}} \equiv \frac{|\theta_{23} - \bar\theta_{23}|}
{(\theta_{23})_{\rm average}}.
\label{eq:bcpt}
\end{equation}
The experimental sensitivity to the possible CPT violation is given 
by the accuracy with which the parameters $a_{\rm{CPT}}$ and/or 
$b_{\rm{CPT}}$ can be measured. In order to estimate the sensitivity
we will treat the neutrino and 
antineutrino channels as different experiments which are 
not combined to fit a common $\Delta m^2_{32}$ and $\theta_{23}$.
In order to establish an effect we therefore need to compare
the values of the parameters $a_{\rm{CPT}}$ and $b_{\rm{CPT}}$, 
which are describing the asymmetry between these two experiments, 
with the corresponding statistical errors of the neutrino 
oscillation parameters determined in \Ref~\cite{Freund:2001ui}, 
\fig~3. Only if the mass squared or mixing angle difference 
between neutrinos and antineutrinos is larger than the respective 
relative
statistical error $\delta \Delta m_{32}^2$ or $\delta \theta_{23}$
of the measurement of $\Delta m_{32}^2$ or $\theta_{23}$, CPT
violation will be detectable on the respective confidence level of 
the statistical evaluation, \ie, the sensitivities $\delta a_{\rm
CPT}$ and $\delta b_{\rm CPT}$ to the CPT violating parameters $a_{\rm
CPT}$ and $b_{\rm CPT}$ are given by:
\begin{subequations}
\begin{eqnarray}
 \delta a_{\rm{CPT}} &\sim& \frac{\delta \Delta m_{32}^2}{2}, \label{eq:arel}\\
 \delta b_{\rm{CPT}} &\sim& \delta \theta_{23}, \label{eq:brel}
\end{eqnarray}
\label{eq:abrel}
\end{subequations}
where $a_{\rm CPT} \leq \delta a_{\rm CPT}$ and $b_{\rm CPT} \leq
\delta b_{\rm CPT}$.
The factor of two in the first relation comes from the
translation from mass squared differences to masses for a hierarchical
(or inverse hierarchical) mass spectrum in \eq~(\ref{eq:acptrel}). As
an example, a statistical error of $7 \%$ in the determination of $\Delta
m_{32}^2$ would correspond to a mass asymmetry sensitivity between neutrinos
and antineutrinos of $3.5 \%$. The sensitivities described by
\eqs~(\ref{eq:arel}) and (\ref{eq:brel}) are plotted in
\fig~\ref{fig:asymmetry}, where
the sensitivity $\delta a_{\rm CPT}$ to the asymmetry $a_{\rm{CPT}}$
is shown as a function of the luminosity ${\cal L}$ by the solid
curve. The respective statistical errors were calculated for a muon
energy of $E = 50$~GeV and for a baseline length of $L =
3000$~km. Similarly, the dashed curve shows the sensitivity $\delta
b_{\rm CPT}$ to the asymmetry $b_{\rm{CPT}}$, with the statistical 
errors calculated for $E = 50$~GeV and $L = 7000$~km. From these curves, one
can, for example, read off for a $10 \, \mathrm{kt}$ detector and $10^{20}$
stored muons per year during 5 years (\cf, the vertical line in the plot) that
$a_{\rm{CPT}} \lesssim 3.8 \cdot 10^{-3}$ and $b_{\rm{CPT}} \lesssim 4.3 \cdot
10^{-2}$. For the mass difference of neutrino and antineutrino we then obtain
for a hierarchical or inverse hierarchical neutrino mass spectrum $|m_{3} -
\overline{m}_{3}| = a_{\rm CPT} (m_3)_{\rm average} \simeq a_{\rm
CPT} \sqrt{\Delta m_{\rm atm}^2} \lesssim 1.9 \cdot 10^{-4} \, \mathrm{eV}$.

\section{Summary and conclusions}
\label{sec:sc}

CPT is a fundamental symmetry preserved in any Lorentz invariant
local quantum field theory. Especially, the SM is a CPT invariant theory.
However, CPT invariance can be violated in models beyond the
SM, like models with extra dimensions or string theory models. It is important
to note that the expected effects of CPT violation depend on the
assumed model. If the Planck mass is close to the TeV scale, such as
it is for models with large extra dimensions, these effects could be observable
in future experiments. We especially addressed the question of CPT violation by
small neutrino mass or mixing angle differences between neutrinos and
antineutrinos, which could, most plausibly, be generated by a mechanism beyond
the SM.
Furthermore, we investigated the sensitivity of future neutrino factory
experiments to the presumably small mass and mixing angle differences in the
$\nu_\mu$ and $\overline{\nu_\mu}$ disappearance channels. Finally, we have
shown that for the neutrino-antineutrino mass difference in a hierarchical
(or inverse hierarchical) neutrino mass spectrum, the upper bound $$ |m_3 -
\overline{m}_3| \lesssim 1.9 \cdot 10^{-4} \, \mathrm{eV}$$ can be obtained.

\begin{acknowledgments}
Support for this work was provided by the Alexander von Humboldt
Foundation [S.M.B.], the Swedish Foundation for International
Cooperation in Research and Higher Education (STINT) [T.O.], the
Wenner-Gren Foundations [T.O.], and the ``Sonderforschungsbereich 375
f{\"u}r Astro-Teilchenphysik der Deutschen Forschungsgemeinschaft''.
\end{acknowledgments}


\begin{thebibliography}{99}

\bibitem{CPT}
G. L{\"u}ders, Kgl. Danske Videnskab. Selskab. Mat.-Fys. Medd. {\bf
28}, No. 5 (1954); W. Pauli, in {\it Niels Bohr and the Development of
Physics}, edited by W. Pauli, L. Rosenfeld, and V. Weisskopf
(Pergamon, London, 1955), p. 30; J.S. Bell, Proc. R. Soc. London A
{\bf 231}, 479 (1955).

\bibitem{Kost1}
{\it CPT and Lorentz Symmetry}, edited by V.A. Kostelecky (World
Scientific, Singapore, 1999); V.A. Kostelecky, hep-ph/0005280;
D.V. Ahluwalia, Mod. Phys. Lett. A {\bf 13}, 2249 (1998), hep-ph/9807267.

\bibitem{Kost2}
V.A. Kostelecky and R. Potting, Nucl. Phys. B {\bf 359}, 545 (1991),
J.R. Ellis, N.E. Mavromatos, and D.V. Nanopoulos,
Int. J. Mod. Phys. A {\bf 11}, 1489 (1996), hep-th/9212057;
I. Mocioiu, M. Pospelov, and R. Roiban, hep-ph/0108136.

\bibitem{Groom:2000in} 
D.E. Groom {\it et al.} (Particle Data Group), Eur. Phys. J. C {\bf
15}, 1 (2000), {\tt http://pdg.lbl.gov/}.

\bibitem{Angelopoulos:1998dw} 
CPLEAR Collaboration, A. Angelopoulos {\it et al.}, Phys. Lett. B {\bf
444}, 52 (1998).

\bibitem{Zavrtanik:2001zq} 
CPLEAR Collaboration, D. Zavrtanik {\it et al.},
Nucl. Phys. B Proc. Suppl. {\bf 93}, 276 (2001).

\bibitem{Leonidopoulos:2001ci} 
Belle Collaboration, C. Leonidopoulos, hep-ex/0107001.

\bibitem{Geer:1998iz} 
S. Geer, Phys. Rev. D {\bf 57}, 6989 (1998), {\it ibid.} {\bf 59},
039903(E) (1999), hep-ph/9712290.

\bibitem{Blondel:2000gj} 
A. Blondel {\it et al.}, Nucl. Instrum. Meth. A {\bf 451}, 102 (2000).

\bibitem{Barger:1999jj} 
V.D. Barger, S. Geer, R. Raja, and K. Whisnant, Phys. Rev. D {\bf 62},
013004 (2000), hep-ph/9911524.

\bibitem{Freund:1999} 
M. Freund, M. Lindner, S.T. Petcov, and A. Romanino, Nucl. Phys. B
{\bf 578}, 27 (2000), hep-ph/9912457; M. Freund, P. Huber, and M. Lindner,
Nucl. Phys. B {\bf 585}, 105 (2000), hep-ph/0004085.

\bibitem{Cervera:2000kp} 
A. Cervera {\it et al.}, Nucl. Phys. B {\bf 579}, 17 (2000), hep-ph/0002108.

\bibitem{Albright:2000xi} 
C. Albright {\it et al.}, FERMILAB-FN-692, hep-ex/0008064.

\bibitem{Adams:2001tv} 
T. Adams {\it et al.}, hep-ph/0111030.

\bibitem{Freund:2001ui}
M. Freund, P. Huber, and M. Lindner, Nucl. Phys. B {\bf 615}, 331
(2001), hep-ph/0105071.

\bibitem{Coleman:1998ti}
S.R. Coleman and S.L. Glashow, Phys. Rev. D {\bf 59}, 116008 (1999),
hep-ph/9812418.

\bibitem{Barger:2000iv}
V. Barger, S. Pakvasa, T.J. Weiler, and K. Whisnant,
Phys. Rev. Lett. {\bf 85}, 5055 (2000), hep-ph/0005197.

\bibitem{atmosS-K}
Super-Kamiokande Collaboration, Y. Fukuda {\it et al.},
Phys. Rev. Lett. {\bf 81}, 1562 (1998), hep-ex/9807003;
Super-Kamiokande Collaboration, Y. Fukuda {\it et al.},
Phys. Rev. Lett. {\bf 82}, 2644 (1999), hep-ex/9812014;
Super-Kamiokande Collaboration, S. Fukuda {\it et al.},
Phys. Rev. Lett. {\bf 85}, 3999 (2000), hep-ex/0009001;
Super-Kamiokande Collaboration, T. Toshito, hep-ex/0105023.

\bibitem{solar}
B.T. Cleveland {\it et al.}, Astrophys. J. {\bf 496}, 505 (1998);
Kamiokande Collaboration, Y. Fukuda {\it et al.},
Phys. Rev. Lett. {\bf 77}, 1683 (1996);
GALLEX Collaboration, W. Hampel {\it et al.}, Phys. Lett. B {\bf 447},
127 (1999);
SAGE Collaboration, J.N. Abdurashitov {\it et al.}, Phys. Rev. C {\bf 60},
055801 (1999), astro-ph/9907113;
GNO Collaboration, M. Altmann {\it et al.}, Phys. Lett. B {\bf 490},
16 (2000), hep-ex/0006034.

\bibitem{solarS-K}
Super-Kamiokande Collaboration, S. Fukuda {\it et al.},
Phys. Rev. Lett. {\bf 86}, 5651 (2001), hep-ex/0103032;
Super-Kamiokande Collaboration, M.B. Smy, hep-ex/0106064.

\bibitem{Ahmad:2001an} 
SNO collaboration, Q.R. Ahmad {\it et al.}, Phys. Rev. Lett. {\bf 87},
071301 (2001), nucl-ex/0106015.

\bibitem{Bahcall:2001zu}
J.N. Bahcall, M.C. Gonzalez-Garcia, and C. Pe{\~n}a-Garay, JHEP {\bf
08}, 014 (2001), hep-ph/0106258.

\bibitem{LSND}
LSND Collaboration, C. Athanassopoulos {\it et al.},
Phys. Rev. Lett. {\bf 77}, 3082 (1996), nucl-ex/9605003;
LSND Collaboration, C. Athanassopoulos {\it et al.},
Phys. Rev. Lett. {\bf 81}, 1774 (1998), nucl-ex/9709006;
LSND Collaboration, G.B. Mills, Nucl. Phys. B Proc. Suppl. {\bf 91},
198 (2001);
LSND Collaboration, A. Aguilar {\it et al.}, Phys. Rev. D
{\bf 64}, 112007 (2001), hep-ex/0104049.

\bibitem{stancu}
I. Stancu, in {\it Proceedings of the IXth International Workshop on
``Neutrino Telescopes''}, edited by M. Baldo Ceolin (Venice, 2001).

\bibitem{MAINZ}
Mainz Collaboration, C. Weinheimer {\it et al.}, Phys. Lett. B {\bf
460}, 219 (1999); Mainz Collaboration, J. Bonn {\it et al.},
Nucl. Phys. B Proc. Suppl. {\bf 91}, 273 (2001).

\bibitem{TROITSK}
Troitsk Collaboration, V. Lobashev {\it et al.}, Phys. Lett. B {\bf
460}, 227 (1999); Troitsk Collaboration, V. Lobashev {\it et al.},
Nucl. Phys. B Proc. Suppl. {\bf 91}, 280 (2001).

\bibitem{Law:2001sy}
J. Law, R. W. Ollerhead, and J. J. Simpson, in {\it Proceedings of the 19th
International Conference on Neutrino Physics and Astrophysics - Neutrino
2000}, Sudbury, Ontario, Canada, 16-21 June 2000, Nucl. Phys. B
Proc. Suppl. {\bf 91}, 1-537 (2001).

\bibitem{Lindner:2001kd}
M. Lindner and W. Winter, hep-ph/0111263.

\bibitem{seesaw}
M. Gell-Mann, P. Ramond, and R. Slansky, in {\it Supergravity}, edited
by F. van Nieuwenhuizen and D. Freedman (North Holland, Amsterdam,
1979), p. 315;
T. Yanagida, in {\it Proceedings of the Workshop on Unified
Theory and the Baryon Number of the Universe, Tsukuba, 1979} (KEK,
Tsukuba, Japan, 1979);
R.N. Mohapatra and G. Senjanovi{\'c}, Phys. Rev. Lett. {\bf 44}, 912 (1980).

\bibitem{Arkani-Hamed:1998vp}
N. Arkani-Hamed, S. Dimopoulos, G.R. Dvali, and J. March-Russell,
Phys. Rev. D {\bf 65}, 024032 (2002), hep-ph/9811448.

\bibitem{Grossman:1999ra}
Y. Grossman and M. Neubert, Phys. Lett. B {\bf 474}, 361 (2000),
hep-ph/9912408.

\bibitem{Huber:2001ug} 
S.J. Huber and Q. Shafi, Phys. Lett. B {\bf 512}, 365 (2001), hep-ph/0104293.

\bibitem{Barenboim:2001ac}
G. Barenboim, L. Borissov, J. Lykken, and A.Yu. Smirnov,
hep-ph/0108199.

\bibitem{Bilenkii:1998dt}
S.M. Bilenky, C. Giunti, and W. Grimus, Prog. Part. Nucl. Phys. {\bf
43}, 1 (1999), hep-ph/9812360.

\bibitem{Murayama:2000hm}
H. Murayama and T. Yanagida, Phys. Lett. B {\bf 520}, 263 (2001),
hep-ph/0010178.

\bibitem{Bazarko:2000id} 
MiniBooNE Collaboration, A. Bazarko, Nucl. Phys. B Proc. Suppl. {\bf
91}, 210 (2001), hep-ex/0009056.

\bibitem{Piepke:2001tg} 
KamLAND Collaboration, A. Piepke, Nucl. Phys. B Proc. Suppl. {\bf 91},
99 (2001).

\end{thebibliography}
\end{document}